\documentclass[11pt,a4paper]{article}
\usepackage{graphicx}
\usepackage{amsmath}
\usepackage{amssymb}
\usepackage{amsxtra}

\newcommand{\ba}{\begin{eqnarray}}
\newcommand{\ea}{\end{eqnarray}}
\newcommand{\be}{\begin{equation}}
\newcommand{\ee}{\end{equation}}
\newcommand{\ar}{\arrowvert}
\title{Comparing mesons and $W_L W_L$ TeV-resonances}
\author{
Antonio Dobado, Rafael L. Delgado, Felipe J. Llanes-Estrada$^1$~\footnote{Speaker, {\tt fllanes@fis.ucm.es}. Preprint issued as INT-PUB-15-059. }\\
and Domenec Espriu$^2$,\\ 
$^1$ Dept. Fisica Teorica I, Univ. Complutense, 28040 Madrid, and \\
$^2$  Institut de Ciencies del Cosmos (ICCUB),
Marti Franques 1, 08028 Barcelona, Spain.
}

\begin{document}
\maketitle

\begin{abstract}
Tantalizing LHC hints suggest that resonances of the Electroweak Symmetry Breaking Sector might exist at the TeV scale.
We recall a few key meson-meson resonances in the GeV region that could have high-energy analogues which we compare, 
as well as the corresponding unitarized effective theories describing them.
While detailed dynamics may be different, the constraints of unitarity, causality and global-symmetry breaking,
incorporated in the Inverse Amplitude Method, allow to carry some intuition over to the largely unmeasured higher energy domain.
If the 2 TeV ATLAS excess advances one such new resonance, this could indicate an anomalous $q\bar{q}W$ coupling.
\end{abstract}

\section{Non-linear EFT for $W_LW_L$ and $hh$}
\label{sec:Lag}
The Electroweak Symmetry Breaking Sector of the Standard Model (SM) has a low-energy spectrum composed of the longitudinal $W^\pm_L$, $Z_L$ and the Higgs-like $h$ bosons. Various dynamical relations suggest that the longitudinal gauge bosons are a triplet under the custodial $SU(2)_c$, and $h$ is a singlet. 
This is analogous to hadron physics where pions fall in a triplet and the $\eta$ meson is a singlet. The global symmetry breaking pattern,  
$SU(2)\times SU(2) \to SU(2)_c$ is common to the two subfields.

The resulting effective Lagrangian, employing Goldstone bosons $\omega^a\sim W_L,\ Z_L$ as per the Equivalence Theorem 
(valid for energies sufficiently larger than $M_W$, $M_Z$), in the non-linear representation,  is~\cite{Delgado:2013loa,Espriu,Alonso:2014wta},
\ba \label{bosonLagrangian} {\cal L}
& = & \frac{1}{2}\left[1 +2 a \frac{h}{v} +b\left(\frac{h}{v}\right)^2\right]
\partial_\mu \omega^i \partial^\mu
\omega^j\left(\delta_{ij}+\frac{\omega^i\omega^j}{v^2}\right) \nonumber
+\frac{1}{2}\partial_\mu h \partial^\mu h \nonumber  \\
 & + & \frac{4 a_4}{v^4}\partial_\mu \omega^i\partial_\nu \omega^i\partial^\mu
 \omega^j\partial^\nu \omega^j +
\frac{4 a_5}{v^4}\partial_\mu \omega^i\partial^\mu \omega^i\partial_\nu
\omega^j\partial^\nu \omega^j  +\frac{g}{v^4} (\partial_\mu h \partial^\mu h )^2
 \nonumber   \\
 & + & \frac{2 d}{v^4} \partial_\mu h\partial^\mu h\partial_\nu \omega^i
 \partial^\nu\omega^i
+\frac{2 e}{v^4} \partial_\mu h\partial^\nu h\partial^\mu \omega^i
\partial_\nu\omega^i
\ea
 This Lagrangian is adequate to explore the energy region 1-3 TeV $\gg$ 100 GeV, and contains seven parameters. 
Their status is given in~\cite{Delgado:2013loa} and basically amounts to $a\in(0.88,1.3)$ (1 in the SM), $b\in (-1,3)$ (1 in the SM) and the other, NLO parameters (vanishing in the SM) largely unconstrained.
This is a reasonably manageable Lagrangian for LHC exploration of electroweak symmetry breaking in the TeV region, before diving into the space of the fully fledged effective theory~\cite{Alonso:2014wta}.

Partial wave scattering amplitudes in perturbation theory
$A_{I}^J(s)= A^{(LO)}_{IJ}(s)+A^{(NLO)}_{IJ}(s)\dots$   
for $\omega\omega$ and $hh$, have been reported to NLO in~\cite{Delgado:2015kxa}. For example, the LO amplitudes of $I=$ 0, 1 and 2, and the $\omega\omega\to hh$ channel-coupling one are
\begin{eqnarray} \label{LOamps}
A_0^0(s)  =  \frac{1}{16 \pi v^2} (1-a^2) s  \ \ \ \ \ \ \ \ A_1^1(s)  =  \frac{1}{96 \pi v^2} (1-a^2) s  \nonumber \\
A_2^0(s)  =  -\frac{1}{32 \pi v^2} (1-a^2)s \ \ \ \ \ \ \ \ 
M^0 (s)   =  \frac{\sqrt{3}}{32 \pi v^2} (a^2-b)s \nonumber 
\end{eqnarray}
and we see how  any small separation of the parameters from the SM value $a^2=b=1$ leads to energy growth, and eventually to strong interactions.
To NLO, the amplitudes closely resemble those of chiral perturbation theory
\be \label{pertamplitude}
A_{IJ}^{(LO+NLO)}(s) = K s + \left( B(\mu)+D\log\frac{s}{\mu^2}+E\log\frac{-s}{\mu^2}\right) s^2 
\ee
with a left cut carried by the $Ds^2\log s$ term, a right cut in the $Es^2\log (-s)$ term, and the $Ks+Bs^2$ tree-level polynomial. 
$B$, $D$ and $E$ can be found in~\cite{Delgado:2015kxa} and satisfy perturbative renormalizability (in the chiral sense).

\section{Resonances}
The perturbative amplitudes in Eq.~(\ref{pertamplitude}) do not make sense for large $s$ (TeV-region) where they violate unitarity
${\rm Im} A_{IJ} = |A_{IJ}|^2 $, relation satisfied only order by order in perturbation theory, namely ${\rm Im} A^{({\rm NLO})}_{IJ} = 
| A^{({\rm LO})}_{IJ}|^2$. 

In hadron physics, the solution is to construct new amplitudes that satisfy unitarity exactly and reproduce the effective theory at low energy (see the lectures~\cite{Truong:1990du}) via dispersive analysis. 
This combination of dispersion relations with effective theory exploits all model-independent information in the two-body experimental data, and is known in both the electroweak symmetry breaking sector and the QCD sector of the Standard Model~\cite{Dobado:1989gr}. 
A salient example is the NLO Inverse Amplitude Method,  
\begin{equation} \label{IAM}
A_{IJ} = \frac{\left( A^{(LO)}_{IJ}\right)^2}{A^{(LO)}_{IJ}-A^{(NLO)}_{IJ}}
\end{equation}
a simple formula that can be rigorously generalized to two channels of massless particles by upgrading the various $A$ to matrices. The denominator of 
Eq.~(\ref{IAM}) allows for  scattering resonances (poles in the 2nd Riemann sheet).

\begin{figure}
\includegraphics[angle=90,width=0.52\textwidth]{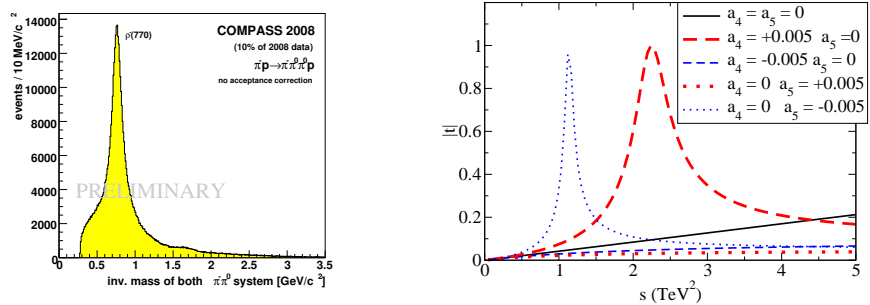}
\includegraphics[width=0.45\textwidth]{FIGS.DIR/ModIAM11.eps}
\caption{Left: the physical $\rho$ in the COMPASS $\pi\pi$ spectrum. (Reprinted from~\cite{Nerling:2010vz}. Copyright 2008, AIP Publishing LLC). Right: a possible equivalent $WW,WZ$ state for various $a_4$, $a_5$. \label{comparerho}}
\end{figure}

In meson physics, the most salient elastic resonance of the $\pi\pi$ system is the isovector $\rho(770)$ meson, that dominates low-energy dipion production in most experiments; for example, its prominence in COMPASS data~\cite{Nerling:2010vz} is visible in the left plot of figure~\ref{comparerho}.
Independently of particular technicolor models, values of $a_4$ and $a_5$ at the $10^{-4}$-$10^{-3}$ level produce a $\rho$-like meson of the electroweak sector in the TeV region. The right panel of figure~\ref{comparerho} demonstrates this.

\begin{figure}
\begin{minipage}{0.48\textwidth}
\includegraphics[width=\textwidth]{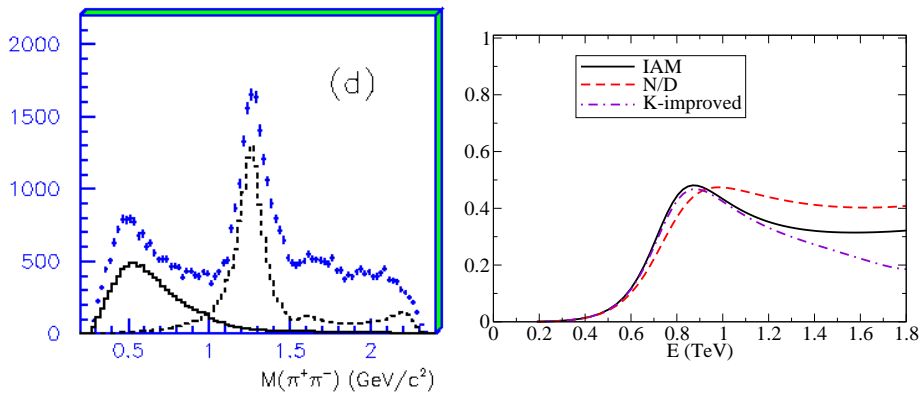} $ $
\end{minipage}
\begin{minipage}{0.48\textwidth}
\includegraphics*[width=\textwidth]{FIGS.DIR/ComparecoupledchannelsJ0.eps}\\
$ $
\end{minipage}
\caption{\label{fig:scalar} 
Left: $\pi\pi$ spectrum with positive charge conjugation
clearly showing an enhancement at low invariant mass, related to the $f_0(500)$ (or $\sigma$) meson;
(Reprinted from~\cite{Bugg:2008eb} with permission. Copyright 2008, AIP Publishing LLC).
Right: $IJ=00$ $\omega\omega$ scattering in the IAM and other unitarization methods producing an equivalent electroweak resonance.}
\end{figure}

The central attraction of the nuclear potential suggested the introduction of a scalar $\sigma$ meson in the $\pi\pi$ spectrum whose existence was long disputed but that is now well established~\cite{Pelaez:2015qba}. 
In addition to detailed dispersive analysis, it gives strength to the low-energy $\pi\pi$ spectrum if the $\rho$ channel is filtered out by cautious quantum number choice, such as $J/\psi\to \omega \pi\pi$ that forces the pion subsystem to have positive charge conjugation because the other two mesons  both have $C=(-1)$. An analysis of BES data by D. Bugg is shown in the left plot of fig.~\ref{fig:scalar}.
The right plot shows the equivalent resonance in $\omega\omega\sim W_L W_L$, that appears for $a\neq 1$ and/or $b\neq a^2$ (if the resonance is induced by $b$ alone it is a pure coupled channel one~\cite{Delgado:2015kxa}, that also has analogues in hadron physics, though less straight-forward ones).

\begin{figure}
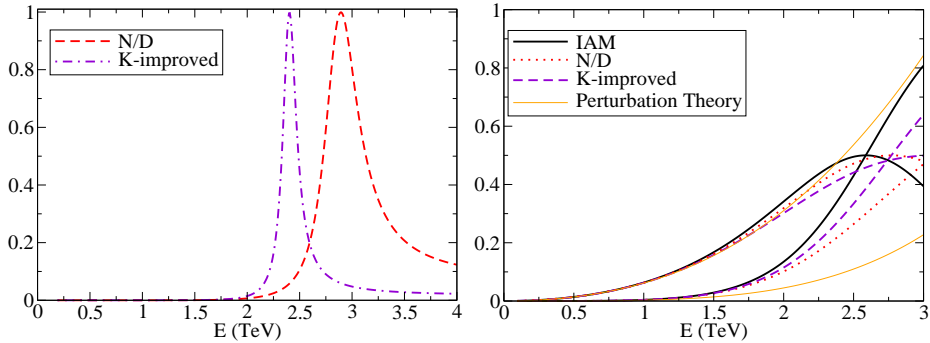

\includegraphics*[width=0.48\textwidth]{FIGS.DIR/ComparecoupledchannelsJ2.eps}\ 
\includegraphics*[width=0.48\textwidth]{FIGS.DIR/CompareI2_a1.15.eps}
\caption{\label{fig:f2} Left: generating an $IJ=02$ resonance in the electroweak sector is possible with adequate values of $a_4$, $a_5$. Right: positive values of $a^2-1$ also generate an isotensor $I=2$ resonance, though this is more disputed~\cite{Espriu}. In hadron physics the isotensor wave is repulsive, and thus, not resonant.}
\end{figure}

The same BES data also reveals another salient meson resonance, the $f_2(1270)$. Partial waves with $J=2$ cannot be treated with the NLO IAM, as 
$A^{2\ \rm LO}_0=0$ but a similar structure has been obtained with the $N/D$ or $K$-matrix unitarization methods, and we show it in figure~\ref{fig:f2}.

\section{ATLAS excess in two-jet events}
\begin{figure}
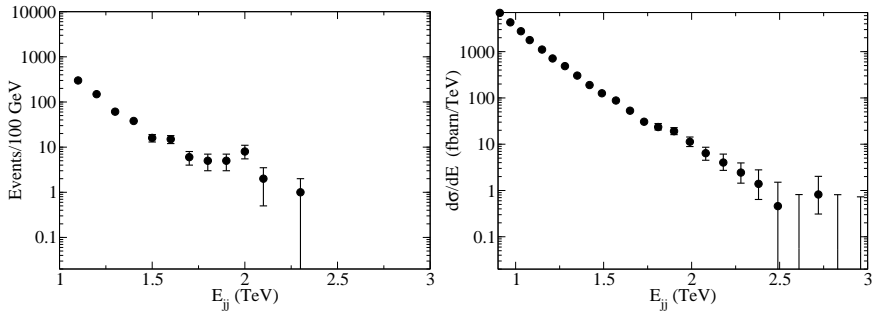

\centerline{\includegraphics[width=0.45\textwidth]{FIGS.DIR/ATLASWZ.eps}\ 
\includegraphics[width=0.45\textwidth]{FIGS.DIR/CMSWW.eps}}
\caption{\label{fig:ATLASdata} Left: replot of the ATLAS data\cite{Aad:2015owa} for $WZ\to 2\ {\rm jet}$, with a slight excess at 2 TeV (also visible in the other isospin combinations $WW$ and $ZZ$, not shown). The jet analysis is under intense scrutiny~\cite{Goncalves:2015yua}.
Right: equivalent CMS data~\cite{Khachatryan:2014hpa} with vector-boson originating jets.  No excess is visible at 2 TeV (though perhaps some near 1.8-1.9 TeV).}
\end{figure}

Renewed interest in TeV-scale resonances is due to a possible excess in ATLAS data~\cite{Aad:2015owa} plotted in figure~\ref{fig:ATLASdata} together with comparable, older CMS data~\cite{Khachatryan:2014hpa} that does not show such an enhancement.
The excess is seen in two-jet events tagged as vector boson pairs by invariant mass reconstruction (82 and 91 GeV respectively). The experimental error makes the identification loose, so that the three-channels cross-feed and we should not take seriously the excess to be seen in all three yet. 
Because $WZ$ is a charged channel, an $I=0$ resonance cannot decay there. Likewise $ZZ$ cannot come from an $I=1$ resonance because the corresponding Clebsch-Gordan coefficient $\langle 1 0 1 0 | 1 0 \rangle$ vanishes. 
A combination of both $I=0,1$ could explain all three signals simultaneously (as would also an isotensor $I=2$ resonance).

A relevant relation imported from hadron physics that the IAM naturally incorporates restricts the width of a vector boson.
This one-channel KSFR relation~\cite{Delgado:2015kxa2} links the mass and width of the vector resonance with the low-energy constants $v$ and $a$ in a quite striking manner,
\begin{equation} \label{KSFR}
\Gamma^{\rm IAM} = \frac{M^3_{\rm IAM}}{96\pi v^2}(1-a^2)\ .
\end{equation}
For $M\sim 2$ TeV and $\Gamma\sim 0.2$ TeV (see fig.~\ref{fig:ATLASdata}), we get $a\sim 0.73$ which is in tension with the ATLAS-deduced bound $a \arrowvert_{2\sigma}>0.88$ at 4-5$\sigma$ level; Eq.~(\ref{KSFR}) predicts that a 2 TeV $J=1$ resonance, with current low-energy constants, needs to have $\Gamma<50$GeV, a fact confirmed by more detailed calculations~\cite{Delgado:2013loa,Espriu}. However scalar resonances tend to be substantially broader.

\begin{figure}
\begin{minipage}{0.45\textwidth}
\includegraphics[width=0.95\textwidth]{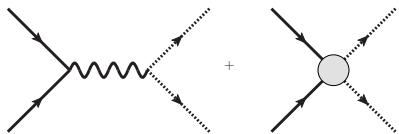}
\end{minipage}
\begin{minipage}{0.5\textwidth}
\caption{Production of a pair of Goldstone bosons by $u\bar{d}$ annihilation through a $W$-meson and
 anomalous BSM vertex enhancing it.\label{fig:Feynman}}
\end{minipage}
\end{figure}

The cross section for the reaction $pp\to W^+Z+X$ for a given $WZ$ Mandelstam $s$, 
and  with the $E^2_{\rm}$ total energy in the proton-proton cm frame,
can be written~\cite{Dobado:2015hha} in standard LO QCD factorization as
\be
\frac{d\sigma}{ds} = \int_0^1 d x_u \int_0^1 dx_{\bar{d}} \delta(s - x_u x_{\bar{d}} E^2_{\rm tot}
f(x_u) f(x_{\bar{d}}) \hat{\sigma}(u\bar{d}\to \omega^+ z) \ .
\ee
The parton-level cross section $\hat{\sigma}$ is calculated, with the help of the factorization theorem, from the effective Lagrangian in Eq.~(\ref{bosonLagrangian}) above. Following~\cite{Dobado:2015hha}, we would
expect an amplitude (from the left diagram of figure~\ref{fig:Feynman})
given by ${\mathcal{M}}=\bar{u} \gamma^\mu_L v (-ig/\sqrt{8})^2 (i/q^2) (k_1-k_2)_\mu$ in perturbation theory. Further, dispersive analysis reveals the need of a vector form factor in the presence
of strong final state rescattering, to guarantee Watson's final state theorem; the phase of the
production amplitude must be equal to that of the elastic $\omega\omega$ scattering amplitude. 
If the later is represented by the Inverse Amplitude Method, the form factor in the $W\omega\omega$ vertex is
$F_V(s)=\left[1-\frac{A_{11}^{(1)}(s)}{A_{11}^{(0)}(s)}\right]^{-1}$.
The resulting cross section~\cite{Dobado:2015hha} was found to be slightly below the CMS bound, and perhaps 
insufficient to explain the possible ATLAS excess. With current precision this statement should not be 
taken to earnestly, but it is nonetheless not too soon to ask ourselves what would happen in the presence of additional 
non-SM fermion couplings. 

Thus, an original contribution of this note is to add to Eq.~(\ref{bosonLagrangian}) a term~\footnote{This is only one of the possible additional operators. 
There is a second one with $R$ fields, and several custodially breaking others. The gauge-invariant version of Eq.~(\ref{newL}) actually modifies the fermion-gauge coupling by a factor $(1+\delta_1)$: this cannot be excluded because it would be the quantity that is actually well measured in $\beta$ decay. The triple gauge boson vertex would then need not coincide with this coupling. However the latter is much less precisely known and there is room for deviations at the 5-10\% level.}
\be \label{newL}
{\mathcal{L}}_{\rm fermion\ anomalous} = \frac{\delta_1}{v^2} \bar{\psi}_L \omega \not \! \partial \omega \psi_L
\ee
(for a derivation see, e.g.~\cite{Bagan:1998vu}). The parton level cross-section is then
\be \label{crosssec}
\left[\frac{d\hat{\sigma}}{d\Omega}\right]_{\rm cm} = \frac{1}{64\pi^2 s} \frac{g^4}{32}\sin^2 \theta \left(
1+\frac{\delta_1 s}{v^2} \right)^2 \ar F_V(s) \ar^2\ ,
\ee
and if $\delta_1\neq 0$ additional production strength appears in the TeV region. The sign of this $\delta_1$ might be determined from the line shape 
due to interference with the background~\cite{Chen:2015cfa}.

\section{Conclusion}

\begin{figure}[tbh]
\begin{minipage}{0.55\textwidth}
\centerline{\includegraphics*[width=0.95\textwidth]{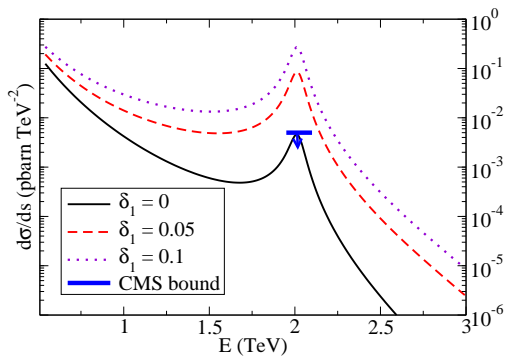}}
\end{minipage}
\begin{minipage}{0.4\textwidth}
\caption{\label{fig:withCMS}  Tree-level $W$ production of $\omega\omega$~\cite{Dobado:2015hha} with final-state resonance; non-zero parameters are $a\!\!=\!\!0.9$, $b\!\!=\!\!a^2$, $a_4\!\!=\!\!7\!\!\times\!\! 10^{-4}$ (at $\mu\!\!=\!\!3$ TeV). Also shown is a CMS cross-section upper bound (see fig.~\ref{fig:ATLASdata}). This can be exceeded with the $\delta_1$ coupling of Eq.~(\ref{newL}).} 
\end{minipage}
\end{figure}

The 13 TeV LHC run II entails larger cross sections and allows addressing the typical $\sigma$, $\rho$-like $\omega\omega$ resonances, at the edge of the run I sensitivity limit as shown in fig.~\ref{fig:withCMS}. The large rate at which such a resonance would have to be produced to explain the ATLAS excess (at the 10fbarn level~\cite{atlasnew}) is a bit puzzling, though it can be incorporated theoretically with the $\delta_1$ parameter.
Hopefully  this ATLAS excess will soon be refuted or confirmed. In any case, the combination of effective theory and unitarity that the IAM encodes is a powerful tool to describe data up to $E=3$TeV  in the electroweak sector if new, strongly interacting phenomena appear, with only few independent parameters. The content of new, Beyond the Standard Model theories, can then be matched onto those parameters for quick tests of their phenomenological viability.

\section*{Acknowledgements}
FLE thanks the organizers of the Bled workshop ``Exploring hadron resonances" for  hospitality and encouragement.
Work  supported by Spanish Excellence Network on Hadronic Physics FIS2014-57026-REDT, and grants UCM:910309, MINECO:FPA2014-53375-C2-1-P, FPA2013-46570, 2014-SGR-104, MDM-2014-0369; its completion was possible at the Institute for Nuclear Theory of the Univ. of Washington, Seattle, with DOE support.


\end{document}